# Framework for Application Mapping over Packet-switched Network of FPGAs : Case studies

Vinay B. Y. Kumar*, Pinalkumar Engineer*, Mandar Datar*,
Yatish Turakhia†, Saurabh Agarwal*, Sanket Diwale‡ and Sachin B. Patkar*

Department of Electrical Engineering, Indian Institute of Technology Bombay, Mumbai, India
Email:*{vinayby, pje, mandardatar, saurabh, patkar}@ee.iitb.ac.in, †yatisht@stanford.edu, ‡sanket.diwale@epfl.ch

*Abstract*—The algorithm-to-hardware High-level synthesis (HLS) tools today are purported to produce hardware comparable in quality to handcrafted designs, particularly with user directive driven or domains specific HLS. However, HLS tools are not readily equipped for when an application/algorithm needs to scale. We present a (work-in-progress) semi-automated framework to map applications over a packet-switched network of modules (single FPGA) and then to seamlessly partition such a network over multiple FPGAs over quasi-serial links. We illustrate the framework through three application case studies: LDPC Decoding, Particle Filter based Object Tracking, and Matrix Vector Multiplication over GF(2). Starting with high-level representations of each case application, we first express them in an intermediate message passing formulation, a model of communicating processing elements. Once the processing elements are identified, these are either handcrafted or realized using HLS. The rest of the flow is automated where the processing elements are plugged on to a configurable network-on-chip (CONNECT) topology of choice, followed by partitioning the 'on-chip' links to work seamlessly across chips/FPGAs.

## I. INTRODUCTION

As applications targeting FPGAs grow more pervasive or when they need to scale, there are matching demands on logic capacity as well as resources such as special-function on-chip resources, I/O and reliable multi-gigabit transceivers. Moore scaling enabled meeting these demands in large part. As with general purpose processors, more than Moore scaling with FPGAs is enabled by multiple FPGA platforms—the classic use-cases of which are ASIC prototyping, Emulation and Hardware-acceleration of applications and also more recently for datacenter applications [1].

Although commercial HLS tools such as Vivado [2]—given good user directives—are capable of producing hardware of quality comparable to handcrafted designs, it is not within the ready scope of HLS tools to address the issue of scalability. This problem becomes even more tricky because of the fragmentation in the ways the multi-FPGA platforms are built, particularly in terms of the variety in the nature of host to FPGA/s and inter FPGA links, and underlying custom interfaces. Dally et. al. [3] recently advocated for design productivity through modular designs with standardized interfaces on a network-on-chip abstraction. In the current context, such a standard interface can abstract the variety in the physical links.

In this work we begin to explore the scalability of applications/algorithms (used interchangeably henceforth)—particularly those amenable to be expressed in a data-flow manner—through a network abstraction, and an automation framework that would simplify exploration of this complex design space in mapping to a given multi-FPGA platform. In particular, we map the application task graph to a packet-switched Network-on-Chip (NoC), and extend the NoC abstraction across FPGAs communicating over quasi-serial links. The path from a higher-level specification of the application to a task-graph with precedence constraints, followed by coarsening and identifying the partition across chips is not discussed in this work (related earlier work: [4]).

We illustrate the framework through three cases studies that could use scalability, each of a different flavor—I. LDPC decoding, min-sum algorithm; II. Particle Filter based Object Tracking; and III. Matrix Vector Multiplication over GF(2). Case I naturally has a message passing structure, unlike II. For case III, although a more straightforward message passing model could have been used, as a way to highlight the role of a domain expert in this step, we use a novel sub-quadratic algorithm by Ryan Williams [5], this incidentally being its first hardware realization. For each case study, in phase-1, we start with a high-level description of the algorithm, express it in message passing formulation, followed by realization of the processing elements either by HLS (Vivado) or custom design. Phase-2 automates the process of integrating these processing elements onto a network-on-chip (NoC) architecture (auto-generated by a NoC Generator, CONNECT [6]), followed by seamlessly (in a manner oblivious to the designer) partitioning the NoC over multiple FPGAs where the NoC links crossing FPGAs are replaced by stitching-in quasi-serial links implemented over FPGA pins. In other words, this work flow expects the algorithm domain expert (software) to help express the original algorithm in a message passing model (phase-1), the rest of the flow is an automation that gives a scaled design over an NoC or multiple FPGAs. Figure 1 outlines the design flow.

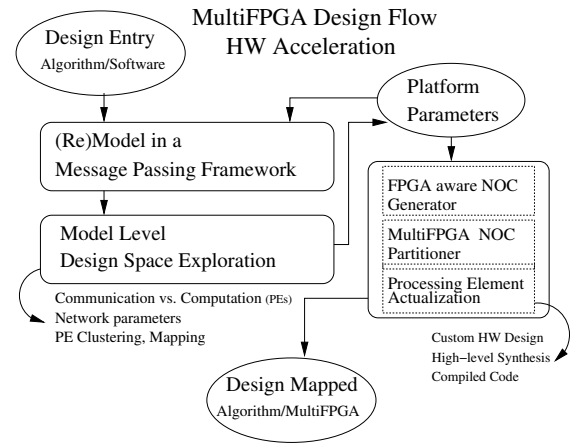

Fig. 1: Design flow: Scaling Hardware Acceleration

This semi-automated framework is a work in progress, and was done with a little manual intervention for the case studies discussed.

*A. Organization*

The rest of the paper is organized as follows. Section II discusses phase-1 of the automation where the algorithm is expressed in a way that helps identification and synthesis of processing elements, followed by wrapping them with suitable adapters before plugging them to CONNECT NoC. Section III, phase-2 of the automation, describes the design of quasi-SERDES endpoints and the automation of partition of the NoC across multiple FPGAs. The next three sections IV, V, and VI discuss the specific case studies mentioned above.

## II. PHASE-1: APPLICATION MAPPING TO NOC

*A. Message passing modeling of the Application*

The algorithm should first be expressed in a message passing formulation. This modeling, at the software level, is best done by the domain expert. The result is a model of software threads—corresponding to processing elements in hardware—communicating in a message passing fashion. For simplicity, we assume the body of the function/thread is executed after all the argument messages on received.

*1) Note on compiler-driven automation:* This phase too can be automated as long as the domain expert annotates the input high-level description appropriately. We have a compiler-driven toy automation flow (Figure 2) for this task, that partitions the Dataflow-Graph (DFG) extracted from a high-level description (straight line code) to be executed on a network of MIPS processors. The DFG parts are compiled to a minimal MIPS instruction set with network-push/pull instructions (FIFO-semantics) added to account for the communication between the DFG parts, taking into account the precedence constraints/schedule. [4] is a follow-up work in this direction focusing on fast scheduling and mapping.

*B. Processing Element Realization and Interfacing to NoC*

The hardware modules corresponding to the nodes of the message-passing graph identified in the previous step could either be designed by hand or a HLS tool. However, at this stage these modules are not yet network/NoC aware. Figure 3 shows the structure of a processing element that makes it pluggable on to an NoC. It consists of three modules: *Data collector*, *Data processing* and *Data distributor*. The *Data*

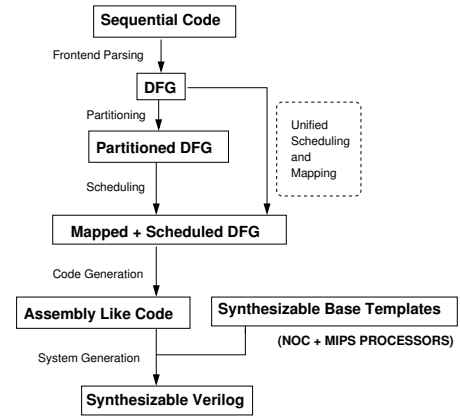

Fig. 2: Basic application partitioning and mapping tool flow

*processing* module is the basic processing element that is synthesized out of the processes/functions from the previous step.

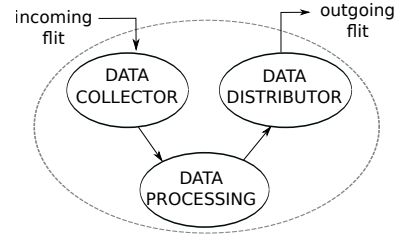

Fig. 3: Structure of Processing elements connecting to NoC

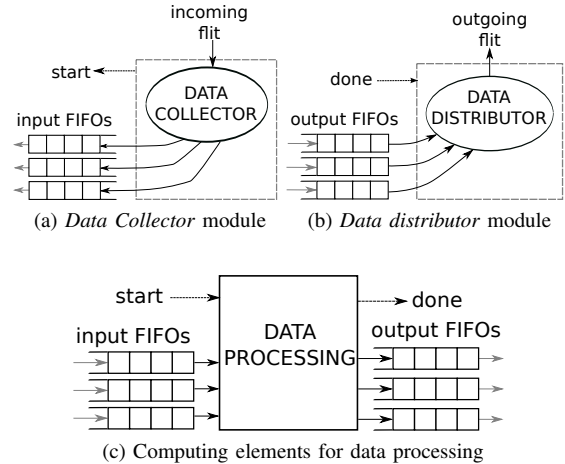

(a) *Data Collector* module  (b) *Data distributor* module

(c) Computing elements for data processing

Fig. 4

*Data collector* and *Data distributor* modules—interfacing an NoC router on one side—are responsible for enabling external communication for processing elements over the NoC. Incoming data (in terms of Flits: basic units of data on NoC links) to the processing element is accepted at the router and processed by the *Data Collector* module, even with the flits arriving in an out-of-order fashion, and is put in appropriate

FIFOs corresponding to the input arguments of the processing element, the *Data processor*. Internal structure of *Data Collector* module is shown in Figure 4a. Once all the data is received and written into FIFOs, *start* is asserted to *Data processing* module. The interface of the *Data processing* module should be as in Figure 4c. Here, as the *start* is asserted, the input data is read from the *input FIFOs*, and once the computation is complete, the results are stored into *output FIFOs* and *done* is asserted. *Data distributor* module, as shown in Figure 4b, prepares the flit data (packet) from results and sends it to network interface of NoC router.

*1) Automation:* As mentioned earlier, the basic processing module could be designed using Verilog HDL or HLS. A script then generates a wrapper around such processing module in form of *Data collector* and *Data distributor* modules. Storage requirements of both input and output memory modules should be known a priori.

## III. PHASE-2: PARTITIONING NoC ACROSS MULTIPLE FPGAS

We use a freely available web-based synthesizable RTL generator for the Network-on-Chip (NoC) infrastructure, named CONNECT (Configurable Network Creation Tool). CONNECT [6] can be used for generating NoCs of arbitrary topology and supports a large variety of router and network configurations. Also, CONNECT incorporates a number of useful features fine-tuned for the FPGA platform.

In extending the NoC links across FPGAs, we require asynchronous links. However, the limited number of pins per FPGA would not support the typical router port-widths and radix counts. We therefore use serializer/deserializer (SERDES) blocks at the interfaces. One would typically use the dedicated multi-gigabit transceiver resources on the FPGA for SERDES links, but for this work, we develop a generic interfacing module that uses the GPIO pins available on any FPGA. As we use more than 1-pin to serialize the flit-transactions across a link (depending on the radix of the router, and the number of pins available), we call them quasi-SERDES.

Assuming an 8-wire physical link, these quasi-SERDES modules (on either side of a link) implement the following protocol—whenever a valid data (valid bit in the flit) in presented as input from router keep it in buffer and start sending 8 bits at a time with MSB first; similarly, whenever a valid 8 bit MSB is received reconstruct output data and put the data on the output port to the router.

Figure 5 shows an example partition of an NoC with four routers on two FPGAs. The router R0 (along with its processing element N0) is mapped onto a separate FPGA. Communication between FPGAs takes place using serializer/deserializer (quasi-SER/DES) links. The processsing elements $N1, 2, 3, 4$ here are as constructed earlier.

### A. Automation

Given an NoC topology and an application mapped to it (as described above), and the decisions (presently user specified) as to 'cuts' that specify a partition on the NoC, an python script automates the process of generating required number of independent parts of the NoC and inserting a pair of quasi-SERDES endpoints on each NoC link cut. The independent part modules of the NoC are synthesized separately and programmed on respective FPGA boards. We have tested this framework between two Altera DE0-Nano boards, as well as two Xilinx Zynq Zedboards (ARM+FPGA).

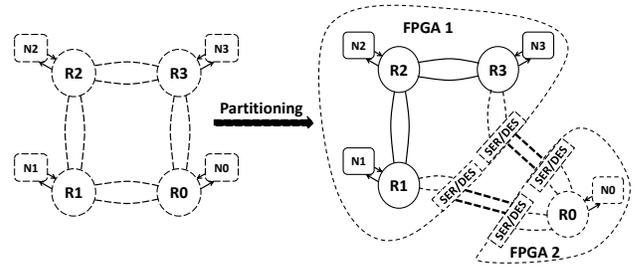

Fig. 5: Example 2-FPGA partition of an NoC

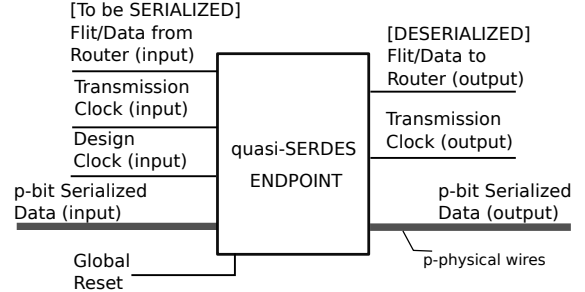

Fig. 6: Quasi-SER/DES Link Endpoint

## IV. CASE STUDY: LDPC DECODING

Listing 1 shows an outline of LDPC decoding based on the popular Min-sum algorithm. Number of data bits, to be decoded is $N$ and $Niters$ is maximum number of iterations for LDPC decoding. Input to LDPC decoder is initial Log-Likelihood Ratio (LLR) of the data. LDPC decoding is done through Check nodes and Bit nodes iteratively, by passing message through dedicated channels between the nodes. Number of channels and interconnection between nodes depends on type of LDPC code. Here, we are using finite projective geometry based LDPC code [7][8] in $GF(2, 2^s)$ with $s = 1$. The message passing model is evident for this application and the processing nodes (the bit and check nodes) are also readily identified.

```
Listing 1: Outline of min-sum LDPC decoding
1  decoded[N] = minsum (data[N], Niter) {
2    do {
3      for (i = 0; i < N; i++) {
4        // Initial LLR values
5        u0(i) = data (i);
6        uij = initial LLRs sent to Check node
7        // j is degree of LDPC nodes
8
9        //Check node processing
10       vij = minimum(uij);
11       //Bit node processing
12       [uij, sum] = sum(vij);
13     }
14   } while(iterations<Niter);
15
16  decoded[N] = sign(sum);
17 }
```

Code listing of check node processing and bit node processing shown in Listing 2 and 3 respectively.

```
Listing 2: Check node processing
1  [v1, v2, v3] = minimum (u1, u2, u3) {
2          v1 = min(u2, u3);
3          v2 = min(u1, u3);
4          v3 = min(u1, u2);
5  }
```

```
Listing 3: Bit node processing
1  [sum, u1, u2, u3] = summation (u0, v1, v2, v3) {
2          sum = u0 + v1 + v2 + v3;
3          u1 = sum − v1;
4          u2 = sum − v2;
5          u3 = sum − v3;
6  }
```

Figures 7 and 8 show typical computing elements for check node and bit node processing respectively.

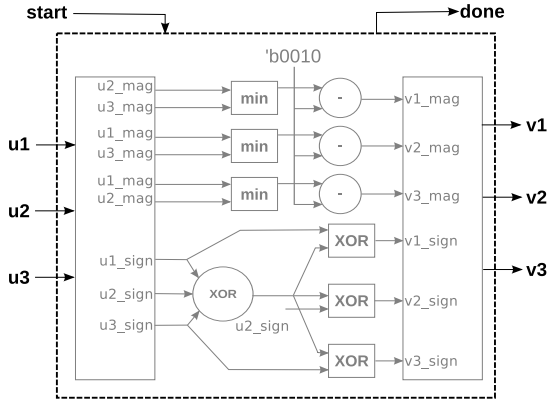

Fig. 7: Check node processing module

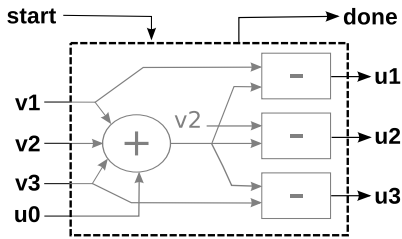

Fig. 8: Bit node processing module

Furthermore, these computing elements have been wrapped with *input FIFOs* and *output FIFOs* for interface compatibility with *Data Collector* and *Data distributor*. The wrappers were generated for both computing nodes, for interfacing them with CONNECT NoC. Table I shows resource utilization of bare computing nodes and computing nodes with wrapper.

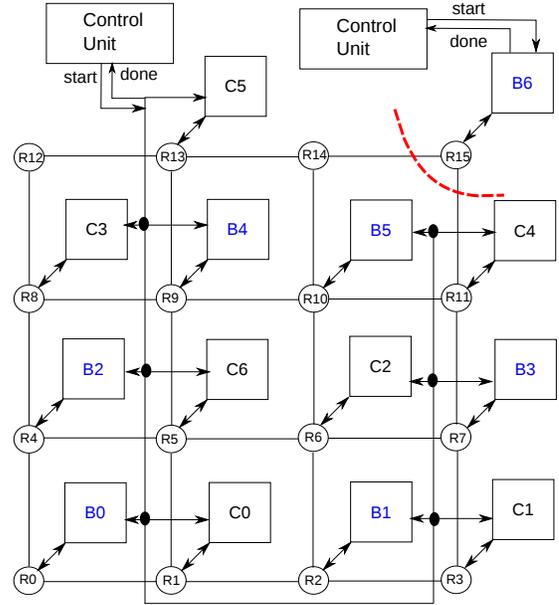

Fig. 9: LDPC decoder using $4 \times 4$ mesh CONNECT NoC

TABLE I: Resource utilization of computing nodes

| Xilinx zc7020 | | Bit node | | Check node | |
|---|---|---|---|---|---|
| | | W/O wrapper | With wrapper | W/O wrapper | With wrapper |
| Resources | Available | Used | Used | Used | Used |
| Slice registers | 106400 | 64 | 297 | 40 | 258 |
| Slice LUTs | 53200 | 110 | 261 | 73 | 199 |

For $N = 7$, both the wrapped computing nodes (bit node and check node), 7 each, are then interfaced to a $4 \times 4$ NoC as shown in Figure 9. Table II shows resource utilization of monolithic LDPC decoder (without NoC, same specs) and same with CONNECT NoC and wrapper. Resource utilization increases mainly due to the NoC being more generic than necessary. Dotted arc in Figure 9 indicates partitioning of NoC for multiple FPGA implementation.

TABLE II: Resource utilization of whole design

| Xilinx zc7020 | | W/O wrapper | | With NoC & wrapper | |
|---|---|---|---|---|---|
| Resources | Available | Used | % | Used | % |
| Slice registers | 106400 | 866 | 1% | 1429 | 1% |
| Slice LUTs | 53200 | 1370 | 2% | 1384 | 2% |

## V. CASE STUDY: PARTICLE FILTER BASED OBJECT TRACKING ALGORITHM

Important steps for our implementation [9] of object tracking based on Sequential Importance Sampling (SIS)

particle filter are listed below:

> **Particle Filtering based object tracking Algorithm:**
> - Calculate **reference histogram**
> - For frames $k \rightarrow 2 \: to \: n$
>   - Initialize $N$ samples $\{x_k^i\}_{i=1..N}$ (Gaussian distribution)
>   - Distance weighted **candidate histograms** for $N$ region of interest (ROI)
>   - Calculate particle weights $\{w_k^i\}_{i=1..N}$ using **Bhattacharya distances** between reference histogram and candidate histograms
>   - New center is estimated using **weighted mean calculation** using centers $\{x_k^i\}_{i=1..N}$ and weights $\{w_k^i\}_{i=1..N}$

Figure 10 shows implementation of particle filter based object tracking algorithm on NoC. For this, we have designed a standalone processing element to compute two important steps—the histogram calculation and calculation of Bhattacharya distances—of particle filter algorithm as shown in Figure 11. Figure 12 show the root node on Node-0 that orchestrates the computations on all other nodes.

Note that this is not necessarily the best way to map this application on an NoC, however, the approach makes exploring variations easier. For instance, the Bhattacharya coefficient calculation block within the current PE could be pulled out and shared as a resource over the network, as a separate processing element.

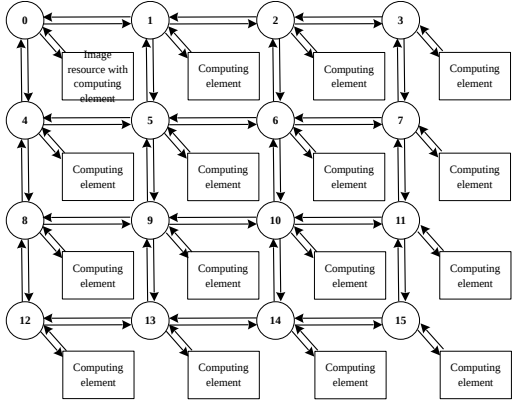

Fig. 10: The Particle filter processing elements mapped over NoC

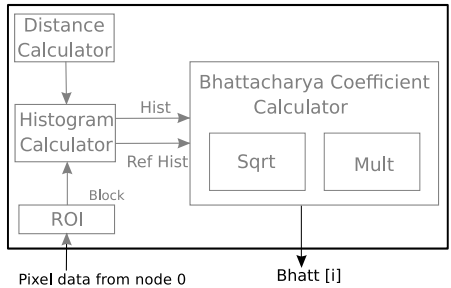

Fig. 11: Compute element for Particle filter based object tracking

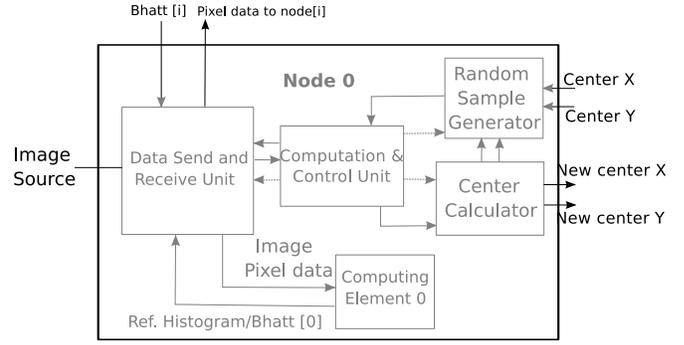

Fig. 12: The processing element on Node 0

Table III shows resource utilization of single processing element without and with wrapper.

TABLE III: Resource utilization of one PE

| Xilinx zc7020 Resources | Available | W/O wrapper Used | % | With NoC & wrapper Used | % |
|---|---|---|---|---|---|
| Slice registers | 106400 | 568 | 1% | 2795 | 2% |
| Slice LUTs | 53200 | 1502 | 2% | 3346 | 2% |
| DSP48E | 220 | 1 | 1% | 20 | 9% |

## VI. CASE STUDY: MATRIX VECTOR MULTIPLICATION OVER GF(2)

Integer factorization is one important application of Matrix Vector Multiplication over GF(2) and solutions have been proposed [10], [11] scaling over about a 1000 chips (ASICs, FPGAs). Block Wiedemann [12] algorithm is often used for this purpose, which needs computations of the form $(AV, A^2V, ..., A^rV)$ involving a very large boolean matrix $A$, and where $V$ has more than one column vectors. Note that $A$ is reused over all the iterations ($r$).

(The sparse floating-point version of the same problem would also have made a good case study, however, over GF(2), the approach used here is particularly communication intensive and through this we also show the impact of the choice of topology.)

### A. Method: Sub-quadratic algorithm to BMVM

Our approach is based on the recently proposed combinatorial algorithm for matrix vector multiplication by Ryan Williams [5]. This approach involves a one-time pre-processing step on $A$, enabling a sub-quadratic time computation of BMVM.

The one-time pre-processing phase involves partitioning the matrix $A$ into tiles of dimensions $k \times k$ as in Figure 13a, followed by construction of $n/k$ look-up tables $\{ LUT_i \mid i : 1 \rightarrow n/k \}$ corresponding to each of the $n/k$ columns of the tiled $A$. $LUT_i$ stores all possible linear combinations of columns of each $k \times k$ tile in the column $i$ of the tiled matrix $A$ (Figure 13a). There can be $2^k$ linear combinations of columns of each $k \times k$ tile, and there are $n/k$ such tiles in a column of $A$. Figure 13b shows the composition of $LUT_i$, which is partitioned into $2^k$ parts, each part storing $n/k$ $k-$bit words such that part$-p$ stores vectors $\{A_{1,i}b_p, A_{2,i}b_p, ..., A_{n/k,i}b_p\}$, where $b_p$ is the $k$-bit vector corresponding to the partition index $p$. In short, the pre-processing step is equivalent to pre-computing and storing all possible products of the tiles of matrix $A$ (ie., $A_{1,1}, A_{1,2}..A_{n/k,n/k}$) with any $k$-bit vector.

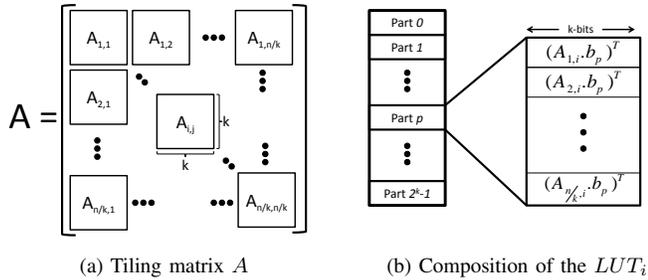

(a) Tiling matrix $A$   (b) Composition of the $LUT_i$

Fig. 13: One-time pre-processing phase

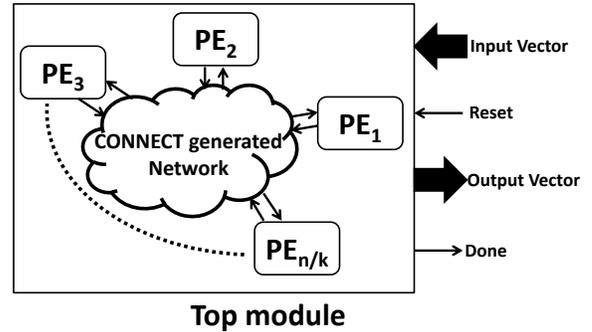

Fig. 14: Top module for Boolean Matrix Vector Multiplication (BMVM).

The computing phase uses this pre-processed information to compute $Av$, for some vector $v$. Let $v$ be likewise partitioned into $n/k$ sub-vectors $(v_1^T, v_2^T, .., v_{n/k}^T)$, and let $v' = Av = (v_1'^T, v_2'^T, .., v_{n/k}'^T)$. For illustration, let $LUT_i$, and $v_i^T$ be with processing node-$i$ (or thread-$i$). As $v_i' = A_{i,1}v_1 \oplus A_{i,2}v_2 \oplus \ldots \oplus A_{i,n/k}v_{n/k}$, if each processing node-$i$ looks-up partition indexed by $v_i$ in $LUT_i$, and send each of the $n/k$ words stored in this partition to the corresponding processing nodes, the result $v_i'$ at each processing node-$i$ is obtained by XOR-accumulating all the incoming $k$-bit messages.

### B. Implementation Details

The one-time precomputed LUTs are mapped to BRAMs on FPGA (Virtex 6 has about $38Mb$). Depending on the problem parameters ($n$ and $k$), not all processing nodes can be mapped to a single FPGA. As per our earlier discussion, we map all the $n/k$ processing elements across all the FPGAs in our NoC-driven multi-FPGA platform. It is important to ensure that while multiple such messages may simultaneously attempt to update a particular product sub-vector $v_i'$, the updates are appropriately serialized to maintain correctness. Since only one flit can be injected and ejected in a single cycle in the NoC, this constraint is automatically ensured. Our implementation uses the following "Network and Router Options" for NoC generated using CONNECT (topology and number of endpoints specified as required):

| Router Type | Simple Input Queued (IQ) |
| ---: | :--- |
| Flow Control Type | Peek Flow Control |
| Flit Data Width | 16 |
| Flit Buffer Depth | 8 |
| Allocator | Separable Input first Round-Robin |

Since number of sub-vectors can be very large ($n/k$), we also implement "folding" (a folding factor $f$), such that a single processing element handles multiple sub-vectors and is provided with a single coalesced look-up table corresponding to the input sub-vectors. We use RIFFA 2.0 [13] to make this acceleration available to the software on the host.

### C. Experimental Results

TABLE IV: Comparative results for $n = 64$ ($64 \times 64$ Matrix) and $k = 8, (fold)f = 2$ (average over 100 experiments). Uses 4 PEs for the hardware and 4 threads for the software version.

| Iterations | Time (in msec) | | Speedup |
| $r$ | Software | Mesh | (over s/w) |
| --- | --- | --- | --- |
| 1 | 0.32 | 0.052 | 6.15 |
| 10 | 1.1 | 0.052 | 21.15 |
| 100 | 5.2 | 0.087 | 59.8 |
| 1000 | 44.2 | 0.58 | 76.2 |

TABLE V: Comparative results for $n = 1024$ ($1024 \times 1024$ Matrix) and $k = 4, (fold)f = 4$ (average over 100 experiments). Uses 64 PEs (and 64 threads for s/w version).

| Iterations | Time (in msec) | | | | |
| | Software | Ring | Mesh | Torus | Fat_tree |
| --- | --- | --- | --- | --- | --- |
| 1 | 4.0 | 0.205 | 0.075 | 0.060 | 0.052 |
| 10 | 22.9 | 1.67 | 0.412 | 0.299 | 0.275 |
| 100 | 204.3 | 16.15 | 3.64 | 2.83 | 2.33 |
| 1000 | 2025.4 | 160.51 | 35.60 | 28.09 | 22.69 |

The evaluation was done on Xilinx Virtex 6 ML605 on an Intel i7 host, hardware-software link between them was implemented using RIFFA 2.0 [13]. The multithreaded message passing software version (processing elements corresponding to threads) was evaluated on a 6 core Xeon (E5-2620). We compare the speed-up from the hardware-software solution compared to this multithreaded pure-software version of the algorithm. The hardware part on the FPGA operates on a 100 MHz clock.

Tables IV and V compare the performance of the multi-threaded message passing software model vs. its equivalent NoC realization on hardware (the times reported for this include the roundtrip time over RIFFA.) In Table V we have evaluated the results for four network topologies implemented on a single FPGA with single cycle hop between adjacent routers, which depict a clear correlation between network cost and performance (the cost increases moving from ring to mesh to torus to fat tree but performance also improves accordingly). When number of iterations are low (1-10), the

27

overheads in terms of host processor - FPGA communication time in hardware and thread *creation/join* time in software, are a dominant component of the overall execution time. For larger iterations (100-1000), the actual computation times dominate and the total execution time increases nearly linearly with number of multiplication iterations.

## VII. CONCLUSION

We presented a semi-automated framework, complementary to existing HLS infrastructure, for scaling algorithms across multiple FPGAs. Through this work-in-progress, we share our experiences evaluating this process with three case studies, each of a different flavor. The application is expressed in the message passing abstraction, and realized over a Network-on-Chip. The network-on-chip abstraction is then extended automatically to seamless work across multiple FPGAs. The proof-of-concept evaluation was done between Xilinx Zynq FPGA (zed)boards.

## VIII. ACKNOWLEDGEMENTS

This work was partially supported by Dr. Suhas Pai (through IITB Heritage Fund) and Nvidia (through CCOE, IITB).